# Aerogel Waveplates

**Pradeep Bhupathi,**[1,*] **Jungseek Hwang,**[1,†] **Rodica M. Martin,**[1] **Jackson Blankstein,**[2] **Lukas Jaworski,**[1] **Norbert Mulders,**[3] **David B. Tanner**[1] **and Yoonseok Lee**[1]

[1]*Department of Physics, University of Florida, Gainesville, Florida 32611-8440, USA*
[2]*Alexander W. Dreyfoos School of the Arts, West Palm Beach, Florida 33401, USA*
[3]*Department of Physics and Astronomy, University of Delaware, Newark, Delaware 19716, USA*
[*]*bhup@phys.ufl.edu*
[†]*jhwang@pusan.ac.kr*

**Abstract:** Optical transmission measurements were made on 98% porosity silica aerogel samples under various degrees of uniaxial strain. Uniaxially compressed aerogels exhibit large birefringence, proportional to the amount of compression, up to the 15% strain studied. The birefringence is mostly reversible and reproducible through multiple compression-decompression cycles. Our study demonstrates that uniaxially strained high porosity aerogels can be used as tunable waveplates in a broad spectral range.

## 1. Introduction

Aerogel, known for more than seventy years [1], is a highly porous material comprised of an entangled network of thin $SiO_2$ strands of 3–5 nm diameter and with an average separation on the order of 10–100 nm [2, 3]. Aerogels can be synthesized with porosity up to 99.9% (only 0.1% silica), making them extraordinarily high in surface area and ultralow in density. Owing to their unique topological nano-porous structure, aerogels have found applications in various areas of industry as super heat insulators [4], light-transmitting fibers [5], non-reflecting materials [6], and transparent walls [7], as well as in academia as Cerenkov counters [8], and for the study of quenched disorder in quantum fluids [9, 10].

Aerogels, as grown, are isotropic on a length scale larger than the structural correlation length ≈ 100 nm. Recently, in the context of investigating the effects of anisotropic disorder in superfluid $^3$He, Vicente et al. [11] suggested that one might induce global anisotropy by uniaxially deforming aerogels and Pollanen et al. [12] then demonstrated global anisotropy in compressed aerogels through optical birefringence measurements. However, no quantitative or spectroscopic information on birefringence was obtained. In this work, we focus on the compression dependent optical properties of 98% porosity aerogel. Specifically, we report optical birefringence, $\Delta n$, of compressed aerogels, observing a difference in the refractive indices along and perpendicular to the axis of compression. Interestingly, the quantity is large enough that compressed aerogels can be used as low-order waveplates in the spectral range from the near infrared to the ultraviolet.

## 2. Principle and experimental method

A uniaxially anisotropic dielectric material exhibits optical birefringence, with different indices of refraction for polarizations parallel and perpendicular to the optic axis. The relative phase shift between light polarized along these two axes is given by

$$\Delta \phi = \frac{2\pi d_c \Delta n}{\lambda} \quad (1)$$

where $d_c$ is the optical path length, $\lambda$ is the wavelength of the light, and $\Delta n = n_e - n_o$ is the birefringence of the material, with $n_e$ and $n_o$ being, respectively, the indices of refraction for the extraordinary and the ordinary rays of the uniaxial medium.

In aerogel, uniaxial strain would modify the distribution of the strand orientation from the presumably isotropic configuration, and consequently introduce global anisotropy. With uniaxially compressed aerogels, however, our analysis must allow for the increase in thickness $d_c$ of the aerogel under compression. If $d_0$ and $L_0$ are the thickness and the length of the uncompressed aerogel, then the thickness of the compressed aerogel is $d_c = d_0(1+ \nu \Delta L/L_0)$ where $\nu$ is the Poisson ratio of the aerogel. For this study we used a previously reported value for 98% aerogel, $\nu = 0.2$ [13].

Our approach, similar to the method used for investigating the birefringence of liquid crystals [14], is shown schematically in Fig. 1. The sample is placed between two polarizing elements, with the first acting as a polarizer and the second as an analyzer. The polarizer is fixed at $-45°$ to the vertical compression axis and the analyzer is adjusted to a specific angle $\theta$ with respect to the polarizer. We then measure the transmitted light intensity as a function of wavelength at various compression rates. Our measurements were performed using a Zeiss MPM800 microscope photometer—a combined microscope and grating spectrometer—equipped with a Xe lamp and a photomultiplier detector. The wavelength range was limited to 330–800 nm by the optical components used. The polarized light beam (≈ 0.5 × 0.3 mm$^2$) was focused on the sample by a 10× condensing lens and the output light was collected by a 10× objective lens located before the analyzer. A sample platform was designed for mounting the aerogel between these lenses so that the light propagation is perpendicular to the axis of compression. A home-made micrometer vise produced compression along the cylindrical axis ranging from 0 to 15% in length. For each compression and decompression, we measured the transmitted light with 4 nm resolution for various angles $\theta$ of the analyzer.

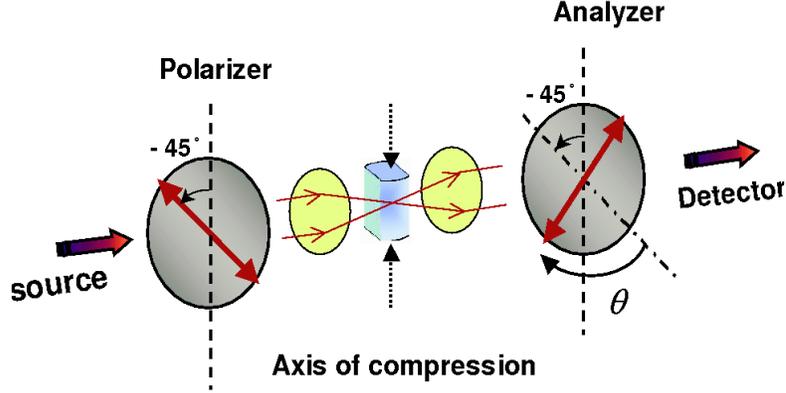

Fig. 1. Schematic diagram of the experimental setup. The axis of compression is fixed in the vertical direction. The polarizer is also fixed at $-45^o$ from the axis of compression. $\theta$ gives the angular position of the analyzer with respect to the polarizer.

With the incident light at $-45^o$, the intensity of the light transmitted by the sample and the analyzer is governed by scattering, absorption, reflection at each interface, and the phase difference between the ordinary and extraordinary rays. Absorption in aerogel for the spectral range of our study is negligible but, as can be seen in our results, Rayleigh scattering needs to be considered, especially at short wavelengths. Therefore, one can express the transmitted intensities $T_\perp$ for perpendicular ($\theta = 90^o$) and $T_{//}$ for parallel ($\theta = 0^o$) orientations of the analyzer to the polarizer as [14, 15]

$$T_\perp = T_o e^{-\frac{r^3 d_c}{\lambda^4}} \sin^2\left(\frac{\pi d_c \Delta n}{\lambda}\right) \qquad (2)$$

$$T_\parallel = T_o e^{-\frac{r^3 d_c}{\lambda^4}} \cos^2\left(\frac{\pi d_c \Delta n}{\lambda}\right) \qquad (3)$$

where $T_o$ represents the wavelength independent loss. The first exponential term is the contribution from Rayleigh scattering, characterized by a dimension $r$. The relative phase retardation, $\Delta\phi$, can be written by combining Eqs. 1–3, to get

$$|\Delta\phi| = \begin{cases} k\pi + 2\tan^{-1}\sqrt{\frac{T_\perp}{T_\parallel}} & k = 0, 2, 4, \ldots \\ (k+1)\pi - 2\tan^{-1}\sqrt{\frac{T_\perp}{T_\parallel}} & k = 1, 3, 5, \ldots \end{cases} \qquad (4)$$

where $k$ is the order of the relative retardation. Clearly, $|\Delta\phi|$ is not uniquely determined. Therefore, when evaluating $|\Delta n|$ from the measurement using Eq. 4, we have to use the proper order of relative retardation $k$ for a given wavelength. Moreover, even the sign of $\Delta\phi$, and hence of $n_e - n_o$, is arbitrary and will have to be obtained from other considerations.

The aerogel samples were synthesized as described in refs. [16] and [17]. The samples, prepared as cylinders, were cut with a high-speed diamond wheel to produce two plane surfaces parallel to the cylindrical axis, in order to eliminate lens effects from the cylinder and to render the optical path constant across the finite-size beam. The dimensions of the samples are listed in Table I along with other parameters obtained from our analysis discussed below.

## 3. Results

Fig. 2 shows the transmittance of sample #1 with $\theta = 90^\circ$ as a function of the wavelength at compressions from 0–15%. These data were taken on the fourth compression-decompression cycle. The reference for all our transmission measurements was a spectrum with parallel polarizer and analyzer ($\theta = 0^\circ$) without sample. Note that an isotropic substance should have zero transmission in the crossed-polarizer configuration, and a sample with $\Delta n \neq 0$ will register a finite transmission except at specific wavelengths where the argument of the sine in Eq. 2 is zero.

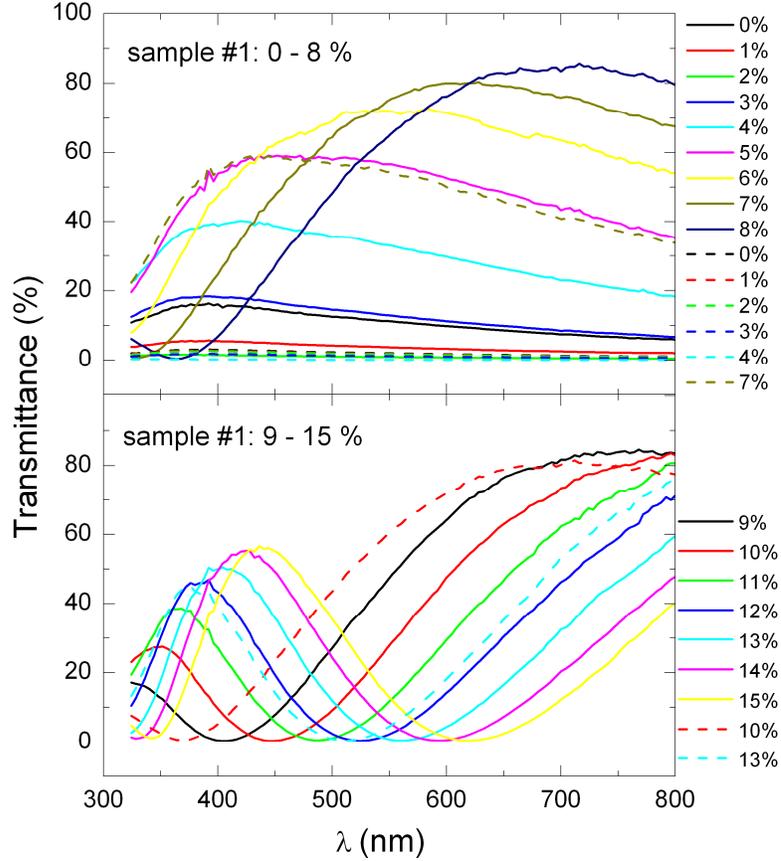

Fig. 2. Transmittance at $\theta = 90^\circ$, $T_\perp$, of sample #1 as a function of wavelength for 0–15% compression. The solid (dotted) traces represent transmittance taken on compression (decompression).

Interestingly, sample #1 shows a small, non-zero transmittance in the uncompressed state indicating some birefringence. We believe that this is due to built-in anisotropy in the sample from the synthesis and some structural damage from previous compression cycles. A substantial amount of anisotropic shrinkage in the drying process has been reported and might be directly related to the built-in anisotropy in uncompressed samples [12]. In sample #1, $T_\perp$ first decreases upon initial compression to reach practically zero at 2% compression. On further compression, the birefringence returns, and an oscillatory behavior develops, as expected from Eq. 2. It seems that a small amount of strain compensates the built-in anisotropy resulting in a sign change in $\Delta n$. Maxima and minima progressively move to longer wavelengths with increasing compression. The dashed lines represent measurements made on the decompression cycle. The existence of hysteresis is manifest. There is about 2% difference in the strain required to achieve similar spectra on compression and decompression. This effect will be discussed in more detail later.

Representative transmission data at $\theta = 0°$, $45°$, and $90°$ for all three samples at zero and maximum compression are shown in Fig. 3. One can clearly see waveplate behavior in the 15% strained sample: the transmittance is independent of analyzer angle ($\theta$) at specific wavelengths, demonstrating that linearly polarized light turns into circularly polarized light. Specifically, sample #1 at 15% compression behaves as a quarter waveplate at wavelengths of 310 nm, 380 nm, 500 nm, and 800 nm and as a half waveplate at 350 nm and 625 nm. These node positions appear periodically in wavenumber [17]. We also show transmission at $25°$ and $70°$ for sample #1 to highlight the waveplate behavior.

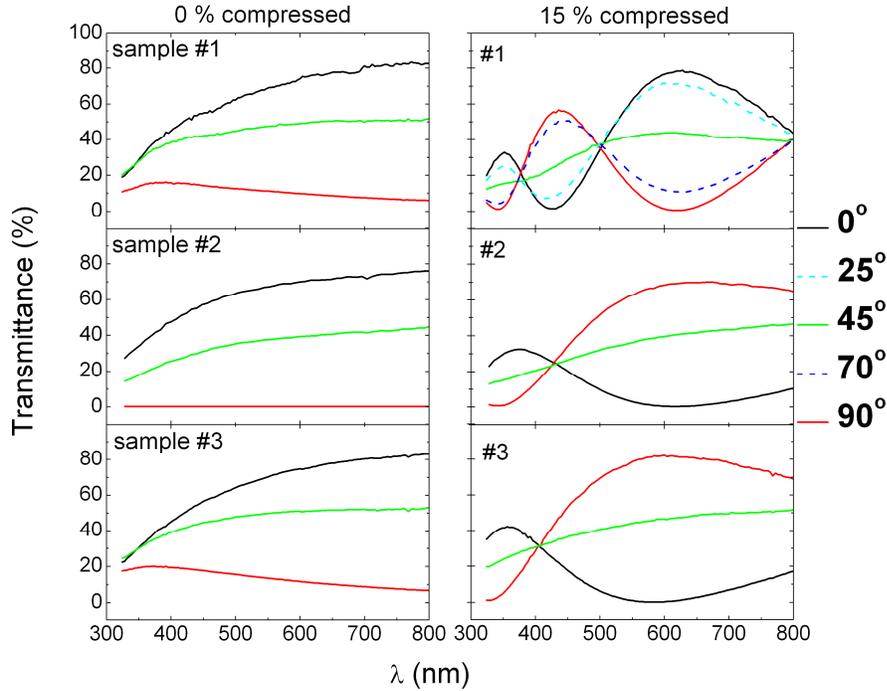

Fig. 3. Transmission of three aerogels at specific angles $\theta$, as a function of wavelength for compressions of 0% (left) and 15% (right). In the top, middle, and bottom rows are data for samples #1, #2, and #3, respectively.

## 4. Birefringence

The birefringence, $\Delta n(\lambda)$, can be extracted in two ways: either from the fit of the transmission data using a model for $n$ and Eqs. 2 and 3 or directly from the transmission ratios based on Eq. 4. Our samples are not optically perfect, suffering from scattering at short wavelengths. The influence of Rayleigh scattering is seen both in the maxima of $T_\perp$ observed under 15% strain (see Fig. 2), which monotonically grow with wavelength, and also in the trace for $\theta = 0°$ at 0% compression in Fig. 3. Rayleigh scattering has been observed previously in an aerogel of a similar porosity [18]. As mentioned above, the effect of scattering is incorporated in Eqs. 2 and 3 with $r$ as the characteristic dimension. The broadband losses are collected into $T_o$. The size of this loss (see Table I) is unexpectedly large considering that material ($SiO_2$) absorption and the mismatch in the index of refraction at the air-aerogel interface are negligible. The loss probably comes from large (larger than the wavelength) scattering centers at the cut surfaces [19]. The final element of our model is the effect of dispersion. We model the birefringence,

$\Delta n$, using a simplified Sellmeier dispersion, or Cauchy formula [15, 20], $\Delta n = B + C/\lambda^2$. The above equations were used to fit our transmission with fitting parameters $r$, $B$, $C$, and $T_o$. The results of our fit are shown as solid lines in the top panel of Fig. 4 along with $T_\perp$ of all our samples at 15% compression. We note that $|\Delta n|$ is determined almost solely from the period of the oscillation in transmission. The actual values of parameters for the fit are listed in Table I.

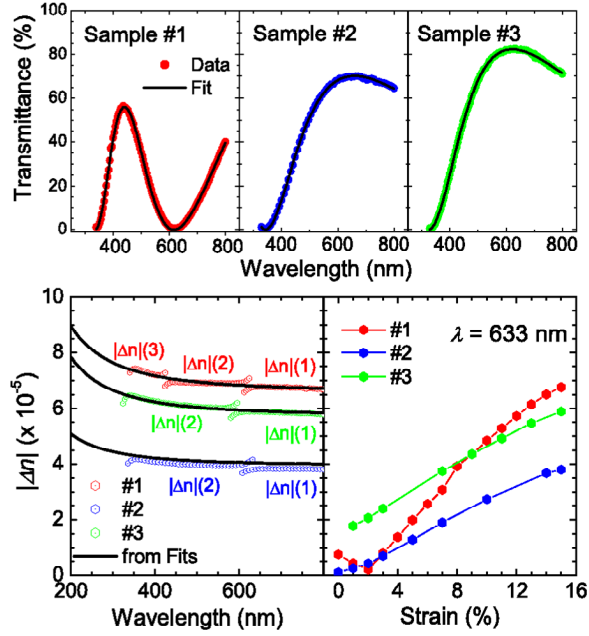

Fig. 4. The top three panels show the measured and fitted $T_\perp$ ($\theta = 90°$) for all three samples at 15% compression. The lower left shows $|\Delta n|$, obtained in two ways (see text) at 15% compression. The lower right panel gives the compression-dependent birefringence, $|\Delta n|$, at 633 nm.

One can directly obtain birefringence using Eq. 4 in which the effects of absorption and scattering are effectively scaled out. However, one needs to specify the proper order $k$ in calculating $|\Delta\phi|$. Each value of $k$ gives a different order of birefringence. We choose the order so as to give a smooth curve for $|\Delta n(\lambda)|$ and to have first order at the longest wavelengths and for a small compression. The results for 15% compression are shown in the lower left panel of Fig. 4. The estimate of $|\Delta n(\lambda)|$ from the fits are also displayed. The agreement is excellent for all three samples and as well for all lower compression values [17]. The dependence of $|\Delta n|$ on strain at 633 nm is shown in the lower right panel. For all samples, $|\Delta n|$ exhibits almost linear dependence on the compression rate.

Table 1. Parameters for aerogel samples

| Sample | $d_o$ (mm) | $L_o$ (mm) | $T_o$ | $r$ (nm) | $B$ | $C$ (nm$^2$) |
|---|---|---|---|---|---|---|
| #1 | 8.78 | 9.48 | 90 | 12.3 | 6.63E–5 | 0.760 |
| #2 | 7.70 | 9.65 | 80 | 13.7 | 3.63E–5 | 0.531 |
| #3 | 4.80 | 7.21 | 80 | 13.5 | 5.53E–5 | 0.854 |

We found that all samples studied did not fully recover to their original lengths after being compressed beyond 5%. Our observation, which will be reported in a separate publication [17], is consistent with the elastic measurements of Gross *et al.* [13]. After 4 cycles of compression to 15%, the total shrinkage amounts to 7–9% for all samples. In particular, the difference in length for sample #1 before and after the third cycle was about 2%. Therefore, the hysteresis in the transmission mentioned earlier (see Fig. 2) can be compensated when the shrinkage is considered in evaluating strain.

## 5. Effective medium approximation

It is a formidable task to establish a microscopic model for our experiment which captures the structural change of the nanometer scale network under uniaxial stress. A sophisticated model should also consider the fractal nature of the aerogel structure. However, one can apply an effective medium approximation (EMA) [21, 22] to our system, recognizing aerogel as a collection of randomly oriented 3–5 nm diameter $SiO_2$ needles, 10–100 nm in length. The effective dielectric constant can be obtained as a solution of the Bruggeman EMA equation

$$f_a \frac{\varepsilon_a - \langle \varepsilon \rangle}{g\varepsilon_a + (1-g)\langle \varepsilon \rangle} + (1-f_a)\frac{\varepsilon_b - \langle \varepsilon \rangle}{g\varepsilon_b + (1-g)\langle \varepsilon \rangle} = 0 \qquad (5)$$

where $f_a$ is the volume fraction occupied by material $a$ and $\varepsilon_a$, $\varepsilon_b$, and $\langle \varepsilon \rangle$ and are respectively the dielectric constants of materials $a$, $b$, and the effective medium. $g$ is a depolarizing factor, related to the shape [23]. Here, we take material $a$ to be $SiO_2$ with $f_a = 0.02$ and $\varepsilon_a = 2.34$; material $b$ to be vacuum with $\varepsilon_b = 1$. We evaluate $\langle \varepsilon \rangle$ for two extreme cases: $g = 0$ and $g = 1/2$. The first is for electric field parallel and the second perpendicular to the $SiO_2$ needles. Then, Eq. 5 gives upper and lower limits for $n = \sqrt{\langle \varepsilon \rangle}$ of $1.0081 < n < 1.0133$. In this model, uniaxial compression would orient the needles preferentially in the plane perpendicular to the compression direction. This leads us to an estimation of the maximum $\Delta n \equiv n_e - n_o \leq -5.2\times 10^{-3}$. Note that the maximum birefringence observed (sample #1 at 15% compression) is $|\Delta n| \approx 7\times 10^{-5}$. The condition for the first order quarter waveplate is $\lambda_q = 4d_c|\Delta n|$ while that for the half waveplate is $\lambda_h = 2d_c|\Delta n|$. From these equations one can see that aerogel can be a tunable waveplate, adjustable for either wavelength or phase shift by uniaxial compression. As an example, sample #1 can be adjusted to give a first order quarter waveplate over any wavelength between 300 nm to 2 $\mu$m.

## 6. Summary

In summary, uniaxially-compressed high-porosity aerogels show large birefringence, proportional to the compression. These compressed aerogels could be used for tunable waveplates over a wide spectral range. Aerogel waveplates might have certain advantages compared to the Babinet compensator: uniform phase retardation in the whole device, low (even first) order waveplates, easy calibration, and multiple adjustable parameters (thickness, compression rate, and porosity).

**Acknowledgements**

We would like to thank Sergei Obukov for useful discussion and acknowledge the support of NSF grants DMR-0239483 and DMR-0803516 (YL), and DOE grant DE-FG02-02ER45984 (DT).